\documentclass[twocolumn,10pt]{asme2e}

\usepackage{graphicx,dblfloatfix}
\usepackage{amsmath}
\usepackage{amssymb}

\usepackage{algorithm}
\usepackage{algorithmic}
\usepackage{multirow}

\confshortname{ISPS 2013}
\conffullname{the ASME 2013 Conference on information storage and processing systems} % \&\\
\confdate{June 24-25}
\confyear{2013}
\confcity{Santa Clara, CA}
\confcountry{USA}

\papernum{ISPS2013-2835}

\title{Embedding Noise Prediction into List-Viterbi Decoding using  Error Detection Codes for Magnetic tape systems}

\author{Suayb S. Arslan  \thanks{Corresponding author. All the
authors are affiliated with Advanced Development Laboratory at Quantum Corporation,  141 Innovation Drive, Irvine,
CA 92617, USA. This work  was partially presented in IEEE/AIP 12th Joint MMM/Intermag conference, Chicago, USA, Jan. 2013.  }
    \affiliation{
	Quantum Corporation\\
	Irvine, California 92617\\
    Suayb.Arslan@Quantum.com
    }	
}

\author{Jaewook Lee
    \affiliation{
	Quantum Corporation\\
	Irvine, California 92617\\
	Jaewook.Lee@Quantum.com
    }
}

\author{Turguy Goker
    \affiliation{
	Quantum Corporation\\
	Irvine, California 92617\\
	Turguy.Goker@Quantum.com
    }
}

\begin{document}

\maketitle

%%%%%%%%%%%%%%%%%%%%%%%%%%%%%%%%%%%%%%%%%%%%%%%%%%%%%%%%%%%%%%%%%%%%%%
\begin{abstract}
{\it A List--Viterbi detector produces a rank ordered list of the $N$ globally best candidates in a trellis search.
A List--Viterbi detector structure is proposed that incorporates the noise prediction  with periodic state-metric
updates based on outer error detection codes (EDCs). More specifically, a periodic decision making process is utilized for a non-overlapping sliding windows of
$P$ bits based on the use of outer EDCs.  In a number of magnetic recording applications, Error Correction Coding (ECC) is adversely effected
by the presence of long and dominant error events.
 Unlike the conventional post processing methods that are usually tailored to a
specific set of dominant error events  or the joint modulation code trellis architectures that are operating on larger state spaces at the expense of increased implementation complexity,
the proposed detector does not use any a priori information about the error event
distributions and operates at reduced state trellis. We present pre-ECC bit error rate performance as well as
%We argue that the complexity of the proposed listing can be reduced using a serial approach that iteratively
%produces $i$-th best path based on the previous found $i-1$ best candidates.
%We further observed that proposed algorithm does not introduce new error events.
the post-ECC codeword failure rates
of the proposed detector using perfect detection scenario as well as practical detection codes as the EDCs are not essential to the overall design.
 Furthermore, it is observed that proposed algorithm does not introduce new error events.  Simulation results show that the proposed algorithm gives improved
 bit error and post-ECC  codeword failure rates
at the expense of some increase in complexity.}
\end{abstract}

%%%%%%%%%%%%%%%%%%%%%%%%%%%%%%%%%%%%%%%%%%%%%%%%%%%%%%%%%%%%%%%%%%%%%%
%\begin{nomenclature}
%\entry{A}{You may include nomenclature here.}
%\entry{$\alpha$}{There are two arguments for each entry of the nomemclature environment, the symbol and the definition.}
%\end{nomenclature}

%%%%%%%%%%%%%%%%%%%%%%%%%%%%%%%%%%%%%%%%%%%%%%%%%%%%%%%%%%%%%%%%%%%%%%
\section*{INTRODUCTION}

At high recording densities, generalized partial response
polynomials with real coefficients are shown to outperform  Partial Response 4 (PR4) and Extended PR4 (EPR4)
detectors. In particular, the idea of noise prediction is embedded into Viterbi decoding in \cite{Evan} and has later become the de facto standard for many recording devices. More specifically, a Noise Predictive Maximum Likelihood (NPML)
detection is proposed which uses a generalized partial response channel with a
polynomial of the form $G(D) = (1-D^2)P(D)$ where $P(D) = 1 + p_1 D
+ p_2 D^2 + \dots + p_L D^L$ is the transfer polynomial and $L$ is the order of the noise whitening filter. Such a finite impulse response filter is
used to approximately whiten the noise at the input of the detector. Introduction of a whitening filter
increases the inter-symbol interference,  the number of trellis states of the system and hence the complexity of the decoding process. However, a feedback loop can be used to reduce the complexity of the detection algorithm at the expense of slight loss in performance due to relying on the past decisions obtained from the trellis.

 \begin{figure*}[htp!]
\centering
\includegraphics[angle=0,height=40mm, width=185mm]{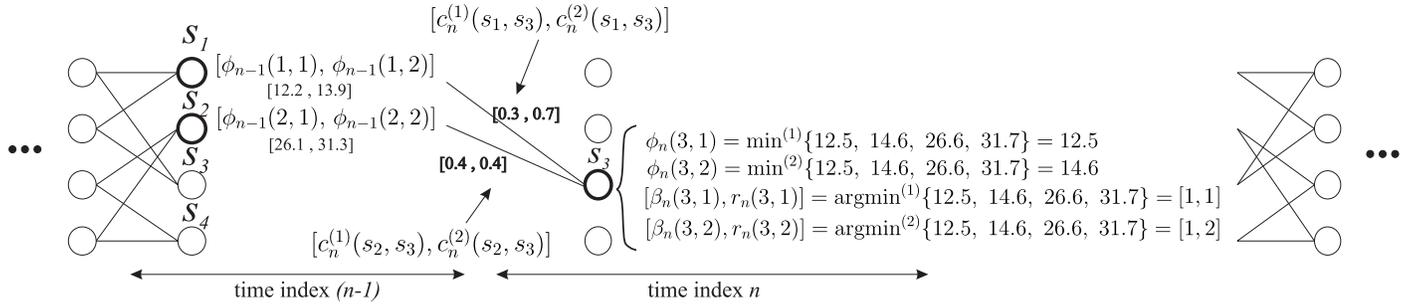}
\caption{Computation of the List-Viterbi metric computations. We assumed 4-state trellis and considered only two significant (most likely) paths ($N=2$) arriving each state of the trellis. The figure shows just one recursion step at time $n$ and the
computation of various quantities in order for the algorithm to
continue.}\label{fig:compute}
\end{figure*}

The performance of the NPML detection must be improved for a reliable
operation of magnetic recording systems, particularly at higher user
densities. There have been two major approaches in the past to improve the detector
performance. One of them focused on combined detector-modulation code \cite{MTR} and the noise predictive decoding architectures using joint trellises
at the expense of larger number of trellis states and hence increased implementation complexity. Joint-trellis
idea has been shown for some of the well known constrained codes and extended PR4 channels to improve overall performance \cite{Siegel}. However, an extension to a specific type of modulation code (such as a twins constrained Maximum Transition Run (MTR) code \cite{Cideciyan2}) is not straightforward due to excessive number of states of the code
that define the constraint. Although such schemes have been shown
to improve performance, they are not used in tape systems mostly due to their poor performances in bursty-error
scenarios.
The other approach was to take the detector structure for granted and
devise post-processing algorithms \cite{Cideciyan1} to improve the performance by eliminating some of the dominant error events at the output of the detector. Such error events are
determined by the recording channel that is disturbed by  various types of noise sources due to mechanical components and media. They are shown to be helpful when the frequency of error event occurrences at the output of the detector is uneven and known to the post processor prior to its operation. Although, post-processing methods are shown to be low complexity, they often result in suboptimum solutions and are not robust i.e., they may attempt to correct some of the dominant error events at the cost of leading to other error events that were originally not part of the detector output. Considering the pros and cons of both approaches in this study, we have developed a List-Viterbi decoding methodology that embedded a noise prediction into the detector and shown considerable BER gains for  generalized partial response channels \cite{Suayb}. Furthermore, the proposed scheme is shown to be robust to miscorrections due to a verification stage using error detection codes.

In conventional tape recording systems,  Reed Solomon (RS) codes are utilized as the error correcting code to recover residual bit errors after the detector. Since the errors at the output of the detector are correlated, different approaches are taken to estimate the post-ECC performance such as multinomial and Block Multinomial Models (BMMs) \cite{Keirn}. This study also presents the semi-analytical post-ECC performance of the proposed detector based on BMMs. We have seen that the proposed List-Viterbi decoding with noise prediction in branch metric computations might be a very important candidate for magnetic tape recording systems giving improved Bit Error Rate (BER) and post-ECC performances
at the expense of some increase in complexity.

%%%%%%%%%%%%%%%%%%%%%%%%%%%%%%%%%%%%%%%%%%%%%%%%%%%%%%%%%%%%%%%%%%%%%%
\section*{THE SYSTEM  MODEL, PROPOSED DETECTOR AND THE POST-ECC PERFORMANCE}

\subsection*{The system model}

Raw user data bits are encoded using a RS code
over $GF(256)$, which can correct up to $t$ bytes of error.
RS codewords are then byte-interleaved, encoded
by a precoder and  a Run Length Limited (RLL) code which satisfies the  run length requirements of 1s and 0s.
After RLL encoding, codewords go through an EDC parity
insertion stage without compromising the RLL constraints. We divide the bit stream into equal size non-overlapping windows and for every
$P$-bit window, equal amount of EDC bits are computed and appended at the end of each window. The unconstrained positions of the RLL code are used to insert the
 EDC parity bits into the bit stram. In this paper, $P$ is called the period
of the proposed algorithm and chosen to be a multiple of
RLL codeword length. The data is finally mapped onto the
symbol sequence $\in \{+1,-1\}$ and written on a storage medium
for readback. Read signal waveform goes through a Low Pass Filter
(LPF), PR4 equalization (symbols $\in \{+2, 0, -2\}$), proposed detector, inverse precoder and RLL/ECC decoding to be able to recover the data.

\subsection*{The proposed detection algorithm}

The proposed algorithm is a combination of periodic updates and a
parallel noise predictive List-Viterbi detection that produces a rank ordered list of the $N$ globally best candidates. The detector starts decoding from the first chunk of the encoded data stream. Using feedbacks from
different path memories, it
decodes the corresponding incoming symbol sequence  and sends the possible candidate
paths to the update stage. Before giving an example, let us provide first the notation we use.

Let us define $\phi_n(j,l), 1\leq l
\leq L+2$ to be the $l$-th lowest accumulated metric to reach state
$j$ ($s_j$) at time $n$ from some starting state at time 0. At time
$n$, we use $\beta_n(j,l)$ to denote the state covered by the $l$-th
best path at time $n-1$, which passes through state $j$ at time $n$.
Similarly, $r_n(j,l)$ characterizes the ranking of the $l$-th best
path at time $n-1$, when this path passes through state $j$ at time
$n$. We will denote the
incremental branch metric (cost) that corresponds to a state
transition $s_j \rightarrow s_k$ (using the $l$-th best path of
$s_j$) at time $n$ using the notation $c_n^{(l)}(s_j,s_k)$.

An example is shown in FIGURE \ref{fig:compute} to explain how the
algorithm computes the accumulated metrics in each time step.  At
time $n$, we would like to compute the best two accumulated metrics
of $s_3$. Based on the trellis structure, we can see that $s_1$ and
$s_2$ make connections with $s_3$ with accumulated metrics
$[\phi_{n-1}(1,1)=12.2, \phi_{n-1}(1,2)=13.9]$ and
$[\phi_{n-1}(2,1)=26.1, \phi_{n-1}(2,2)=31.3]$, respectively.
Considering the different branch metrics corresponding to the state
transitions $s_1 \rightarrow s_3$ and $s_1 \rightarrow s_3$, it is
easy to see that we will end up with four possible path metrics
arriving $s_3$: \{12.5, 14.6, 26.6, 31.7\}. We retain the smallest
two in our algorithm i.e., $\phi_{n}(3,1)=12.5$ and $\phi_{n}(3,2)=14.6$ to be the
accumulated metrics of $s_3$ at time $n$ and discard the other
paths. Note that both of these survival paths pass through state
$s_1$ (i.e., $\beta_n(3,1) = 1$ and $\beta_n(3,2) = 1$) using the
first and the second best path of $s_1$, respectively (i.e.,
$r_n(3,1) = 1$ and $r_n(3,2) = 2$).

After the computation of accumulated metrics as in the example, the detector generates the corresponding incoming symbol sequences and send the possible candidate
paths to the update stage. The update stage periodically  makes a
decision on the correct path  and updates the accumulated metrics. After making a decision
on a particular path using the outcome of EDC decoding,
updated accumulated metrics are forwarded to the detector for
further processing i.e., they are used as the initial conditions for decoding the next window of $P$ bits.
 We repeat the same set of operations for the
decoding of each chunk. Since a decision is made after each update
step, the algorithm continually outputs the decoded bits. For more technical details, we refer the reader to \cite{Suayb}.

\subsection*{The post-ECC performance}

The data recovery performance of modern tape storage systems is usually measured by the post-ECC failure rates.
 In our configuration, an  RS code encodes a data
block of size $m_b = 1960$ bits into $245 + 2t$ ECC symbols,
where $t=5$ is the RS code correction power. In other words, the ECC decoder can correct any combination of $T \leq t = 5$ symbol errors. The Codeword Failure Rate (CFR) is defined as the
probability of reading a codeword not correctable by the RS decoder
and related to Hard Bit
Error Rate (HBER) via the expression HBER = CFR/$m_b$, where $m_b$ is the codeword length in bits. Since the conventional ECC decoders might have to bring the CFR down to $\approx 10^{-9}\sim10^{-13}$, it is infeasible to simulate data to get to these error rates. In this
study, we use the BMM for finding post-ECC CFR
performance \cite{Keirn} particularly for low CFR values. Most disk or tape drive manufacturers report the HBER  to their customers rather than the raw BER
at the output of the detector.  Therefore, it is reasonable to
present the Post-ECC performance in terms of the
CFR.

In that model, we divide the codewords into $M$-symbol equal size blocks. Let the
weight distribution function be $Y(D) = y_1 D + y_2 D^2 + \dots +
y_M D^M$, where $y_w$ is the probability of receiving an $M$-symbol
block, with $w$ errors at the output of the RLL decoder. In our study,
$y_w$s are estimated by way of error counting through simulations.
Once we estimate the probabilities $\{y_i\}_{i=1}^M$ for a given $M$, it is straightforward to
compute the approximate CFRs using the \textbf{Algorithm
\ref{Alg:SRR}} \cite{Keirn}.

%%%%%%%%%%%%%%%%%%%%%%%%%%%%%%%%%%%%%%%%%%%%%%%%%%%%%%%%%%%%%%%%%%%%%%

\begin{algorithm}[b!]
\caption{CFR estimation using\emph{ Block Multinomial
}Model \cite{Keirn}}
\begin{algorithmic}
\STATE \textbf{Initialization:} $j = 1$,   $R(D) = r_1 D + r_2 D^2 +
\dots + r_M D^M = Y(D)$ \STATE \textbf{Recursion:} \FOR{$1 < j\leq
j_{max}$}  \STATE $y_{j,t+1} = \sum_{i \geq t+1} r_i$,  $R(D) :=
R(D) \ast Y(D)$ \ENDFOR \STATE \textbf{Result}: we get: $CFR =
\sum_{j \geq 1}^{j_{max}} \binom{n/M}{j} \times y_0^{n/M-j} \times
y_{j,t+1} $ where $\ast$ denotes polynomial multiplication, $y_0 = 1
- \sum_{j=1}^M y_j$ and $j_{max} = 2t$ is called the truncation
parameter.
\end{algorithmic} \label{Alg:SRR}
\end{algorithm}

%%%%%%%%%%%%%%%%%%%%%%%%%%%%%%%%%%%%%%%%%%%%%%%%%%%%%%%%%%%%%%%%%%%%%%
%%%%%%%%%%%%%%% begin table   %%%%%%%%%%%%%%%%%%%%%%%%%%
%\begin{table}[t]
%\caption{THE TABLE CAPTION USES CAPITAL LETTERS, TOO.}
%\begin{center}
%\label{table_ASME}
%\begin{tabular}{c l l}
%& & \\ % put some space after the caption
%\hline
%Example & Time & Cost \\
%\hline
%1 & 12.5 & \$1,000 \\
%2 & 24 & \$2,000 \\
%\hline
%\end{tabular}
%\end{center}
%\end{table}
%%%%%%%%%%%%%%%% end table %%%%%%%%%%%%%%%%%%%
%%%%%%%%%%%%%%%%%%%%%%%%%%%%%%%%%%%%%%%%%%%%%%%%%%%%%%%%%%%%%%%%%%%%%%

\begin{figure}[t!]
\centering
\includegraphics[height = 59mm, width=\columnwidth]{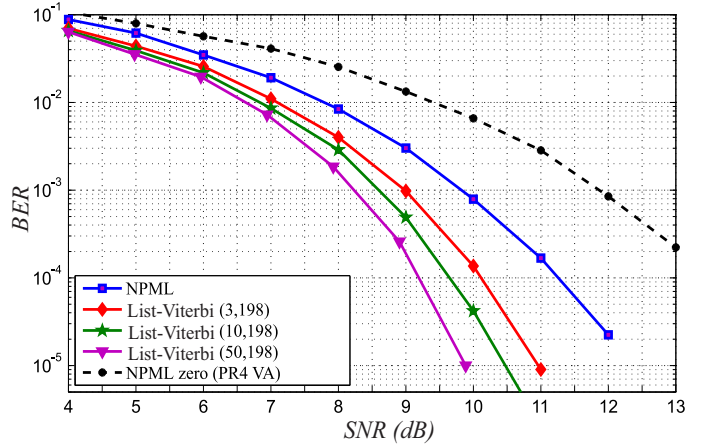}
\caption{Simulation result using Lorentzian channel model using
$D_c=3.25$, $\beta=0.5$ and perfect error detection. LP: Linear
prediction}\label{fig:sim1}
\end{figure}

%%%%%%%%%%%%%%%%%%%%%%%%%%%%%%%%%%%%%%%%%%%%%%%%%%%%%%%%%%%%%%%%%%%%%%
\section*{NUMERICAL RESULTS}

We assumed PR4 signalling, set the order of noise whitening filter
$L=3$ and considered various values of $N$ in all the simulations. The trellis has four states and there are three bits in feedback loop.  Our channel model is the first order position
jitter model \cite{moon2} based on a Lorentzian step response. Electronics and stationary transition
noise samples are added to the signal waveform at the input of the
LPF as modeled in \cite{Cideciyan2}, i.e., as a mixture of the electronics and transition noises. SNR is computed at the input of the LPF and given
by $2/(N_0 + N_m)$ where $N_0$ is double the two sided
spectral densities of a white Gaussian noise source and $N_m$ is the spectral density of the colored/transition noise. We approximate the ratio of the transition noise
to the total noise power by $\beta = N_m / (N_0 + N_m)$ \cite{Cideciyan2}. We assumed
perfect timing recovery, a $5^{th}$ order Butterworth LPF with a 3dB
cutoff frequency. PR4 equalizer is based on
the minimum mean square error criterion. Whitening filter
coefficients ($p_1, p_2$ and $p_3$) are selected using linear prediction
 in minimum least squares sense based on the channel noise
samples.

\begin{figure}[t!]
\centering
\includegraphics[height = 64mm, width=\columnwidth]{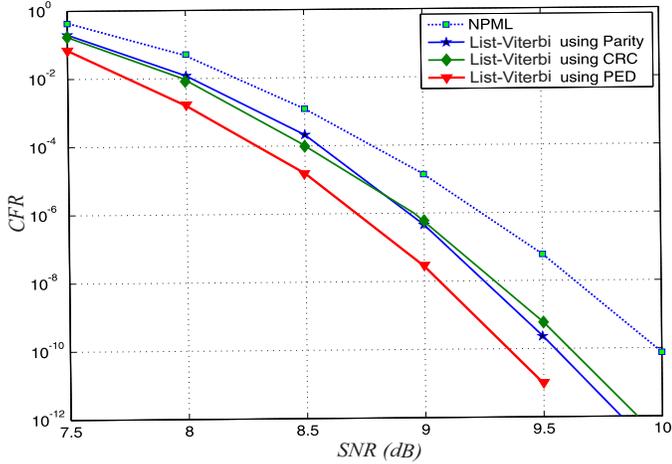}
\caption{CFR using the Lorentzian channel model at $D_c=3.25$ with $\beta=0.5$. We set $N=3$ and $P=198$bits.}\label{fig:simsrr}
\end{figure}

FIGURE \ref{fig:sim1} shows selected results assuming Perfect Error
Detection (PED) for
 $P=198$ bits and various $N$. We also
include the performance of the conventional NPML both for adaptive and fixed
whitener coefficients ($p_1=0, p_2=0, p_3=0$) i.e.,
PR4 Viterbi Algorithm. At a BER of $10^{-4}$, using
$N=50$, a gain of $1.9$dB is observed
over the conventional NPML detector. In a more practical scenario with $N=3$, an average gain of 1dB is observed at
the same operating BER. Also, these performance curves show an
almost 2.5dB gain at a BER of $10^{-5}$ using $N=50$. Those ideal error correction-based
performance curves may serve as benchmarks for the ultimate system
performance. We have also tested our detector combined with actual EDCs such as Cyclic Redundancy Check (CRC) codes. We have seen  the performance degradation is minor relative to
PED case and is only
clear at lower SNR and higher linear densities $D_c = PW50/T$, where $PW50$ is the pulse width measured at half the peak amplitude of the channel's step response and $T$ is the bit period.

Finally, we present approximate post-ECC CFR performances in FIGURE \ref{fig:simsrr}. We use $D_c = 3.25$ for the List-Viterbi detector whereas, in order
for a fair comparison, we assumed that $T$ increases to $T_{NPMLD} = 67\times T/66$ for the NPML detection.  This approximately corresponds to $D_c=3.2$. We also note that SNR loss due to this density increase is minor. We choose $M=17$ bytes for BMM  and use
a $(255,245)$ RS code with $t=5$. The overall 0.85dB
pre-ECC gain at a BER of $10^{-3}$ translates to around 0.65dB post-ECC
gain using PED. This gain is reduced to 0.45dB when the proposed detector is
used with actual detection codes. We note that this gain will increase at higher BER operating points and is
greater than the reported CFR gains in \cite{Keirn} for a single
parity bit post processor per RLL codeword. This basically
shows that although the pre-ECC system performance is slightly
effected by the imperfection of actual detection codes, the post-ECC
performance can severely be effected. On the other hand, with the
provided flexibility, the post-ECC gains can
 be improved by increasing $N$ and/or having smaller $P$ as long as the proposed detector
satisfies the system-specific physical constraints.

\end{document}